Superconducting dome and microstructure properties of $Rb_{0.8}Fe_{1.6+x}Se_2$ superconductors

# Superconducting dome and microstructure properties of $Rb_{0.8}Fe_{1.6+x}Se_2$ superconductors

Zhiwei Wang, Zhen Wang, Yao Cai, Zhen Chen, Chao Ma, Huaixin Yang and Jianqi Li

Beijing National Laboratory for Condensed Matter Physics, Institute of Physics, Chinese Academy of Sciences, Beijing 100190, People's Republic of China

E-mail: LJQ@aphy.iphy.ac.cn

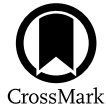

## Abstract

High quality $Rb_{0.8}Fe_{1.6+x}Se_2$ single crystals are grown by a one-step melting growth method. Superconductivity has been observed in all samples with $x \geqslant 0.05$ and the maximum critical temperature $(T_c) \approx 32$ K has been obtained in samples with $0.1 \leqslant x \leqslant 0.2$ from electronic resistivity measurement. A superconducting dome in the electronic phase diagram of $T_c$ as a function of Fe content has been observed for the first time in this $A_xFe_ySe_2$ ($A$ = K, Rb) superconducting system. Two coexisting phases with structural modulation vectors of $q_1 = 1/5$ [$3a^* + b^*$] and $q_2 = 1/2[a^* + b^*]$ have been observed in all superconducting samples by TEM observations. Moreover, phase separation in the superconducting crystals, which can be recognized as micro-stripes in the $a$–$b$ plane, are found to be highly dependent on the Fe content.

(Some figures may appear in colour only in the online journal)

## Introduction

Superconductivity in alkali metal intercalated FeSe-layered compounds $A_xFe_ySe_2$ ($A$ = K, Rb, Cs et al) with $T_c$ above 30 K has inspired great interest in the study of novel superconducting materials [1–5]. In fact, this superconducting system has a defect-driven 122 structure and in general exhibits complex microstructural and magnetic properties resulting from the Fe-deficiency and vacancy ordering [6–8]. Measurements of neutron diffraction and NMR [9] have focused on the co-existence of magnetic order with superconductivity. Optical investigation of $K_xFe_{2-y}Se_2$ superconducting samples has demonstrated the existence of a small energy gap in the superconducting state [10]. ARPES measurements [11–13] reveal the lack of hole pockets in the Fermi surface of the superconducting compounds, making this high $T_c$ superconductivity and the large-moment magnetic order barely explainable by the nesting mechanism. Our previous TEM studies of $K_xFe_{2-y}Se_2$ revealed that these compounds have a rich variety of structural features and a complex microstructure [6, 14, 15]: two coexisting structural phases can be characterized by the modulations of $q_1 = 1/5$ [$a^* + 3b^*$] and $q_2 = 1/2[a^* + b^*]$, which correspond to the antiferromagnetic $K_{0.8}Fe_{1.6}Se_2$ phase and the superconducting phase, respectively. The $q_2$ phase (stripe structure) with a composition of $K_xFe_2Se_2$ is considered to be the superconducting phase that plays a critical role in understanding the essential physical properties of this system. In addition to the notable stripe structure, a nanoscale phase separation also appears in the superconducting samples. The intrinsic phase separation in this type of superconductors has also been observed by many other techniques, such as XRD, ARPES, STEM and so on. For example, Ricci et al reported a nanoscale phase separation in $K_{0.8}Fe_{1.6}Se_2$ and suggested that the coexistence of insulating vacancy-ordered magnetic domains with $\sqrt{5} \times \sqrt{5} \times 1$ structure and metallic non-magnetic domains is responsible for superconductivity [16].

In this paper, we report on the evolution of the superconductivity and microstructural properties of $Rb_{0.8}Fe_{1.6+x}Se_2$ single crystals. A dome-shaped electronic phase diagram has been found for the first time in this kind of FeSe-122 superconductor. In addition, two coexisting phases with modulation vectors of $\mathbf{q}_1 = 1/5[3\mathbf{a}^* + \mathbf{b}^*]$ and $\mathbf{q}_2 = 1/2[\mathbf{a}^* + \mathbf{b}^*]$, as well as phase separation, have been observed in all superconducting samples.

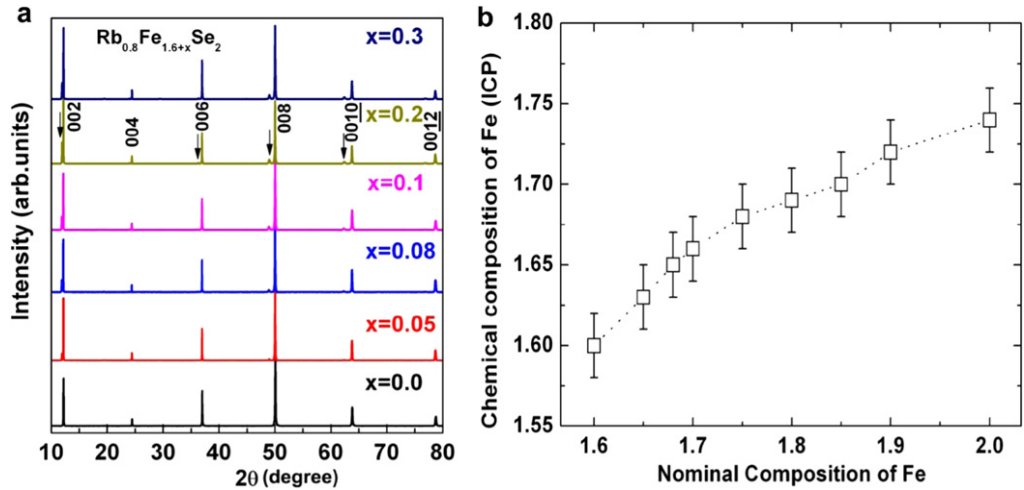

**Figure 1.** (a) XRD patterns of $Rb_{0.8}Fe_{1.6+x}Se_2$ show the coexistence of two structural phases in superconducting materials with the peaks indicated by the arrows. (b) ICP data obtained from $Rb_{0.8}Fe_{1.6+x}Se_2$ single crystals.

## Experimental details

All single crystals of $Rb_{0.8}Fe_{1.6+x}Se_2$ have been grown by a one-step melting growth method. The stoichiometric mixtures of Rb pieces (Alfa, 99.9%), Fe powder (Alfa, 99.9%) and Se pellets (Alfa, 99.999%) with nominal composition were loaded into an $Al_2O_3$ crucible and the crucible was then sealed in an evacuated quartz tube. The tube was slowly heated to 200 °C and maintained at this temperature for 20 h, then heated slowly to 1000 °C and maintained for 5 h, then cooled to 800 °C within 20 h and finally cooled down to room temperature. The obtained crystals have a shiny surface with typical size up to $10 \times 6 \times 1$ mm$^3$.

The temperature dependence of resistivity was measured by a standard four-probe method. The dc magnetic susceptibility was measured with an applied magnetic field of 10 Oe performed on a commercial Quantum Design magnetic property measurement system (MPMS). The SEM images were performed on a Hitachi S-4800 SEM equipped with energy dispersive x-ray (EDX) analysis. X-ray diffraction (XRD) is performed on a Bruker AXS D8 Advanced diffractometer with $2\theta$ from 10° to 80°. The average chemical composition was obtained by inductively coupled plasma atomic emission spectroscopy (ICP-AES), which was performed on a Thermo IRIS Intrepid II. Specimens for the TEM observations were prepared by peeling off a very thin sheet of thickness around several tens of microns from the single crystal. Microstructure analyses were performed on a FEI Tecnai-F20 TEM.

## Results and discussions

XRD and ICP were first performed to characterize the structure and chemical composition of $Rb_{0.8}Fe_{1.6+x}Se_2$ single crystals, respectively. The main diffraction peaks can be well indexed as the known tetragonal layered structure with the lattice parameter of $c = 14.563$ Å, as shown in figure 1(a). It is observed that a series of additional diffraction peaks appear in samples with $x$ more than 0.05, as indicated by arrows in figure 1(a). These peaks, which are very similar to $K_xFe_ySe_2$ samples as reported in [14], are related to the appearance of superconductivity and phase separation. Figure 1(b) shows the average ICP data obtained from a few sets of single crystals. Note that the Fe content could be rather different from the nominal composition for the samples with $x \geqslant 0.1$. On the other hand, Fe content increased continually with the increase of $x$, which is slightly different from that observed in $K_{0.8}Fe_{1.6+x}Se_2$ compounds [14].

Figure 2 shows the temperature dependence of the in-plane resistivity for a series of $Rb_{0.8}Fe_{1.6+x}Se_2$ samples. It is recognizable that the resistivity behaviors obviously depend on the Fe concentration and that superconductivity is observed in samples with $x \geqslant 0.05$. When $x \leqslant 0.05$, resistivity increases with the decrease of temperature, and shows semiconducting behavior, as illustrated in figure 2(a). When $x$ is more than 0.05, all samples show superconductivity with the maximum $T_c$ up to 32 K, as shown in figures 2(b) and (c), where the $T_c$ is chosen as the onset temperature of the downturn where a superconducting transition happens in resistivity curves. Samples with $x$ in the range of $0.1 \leqslant x \leqslant 0.2$ show the highest $T_c$ around 32 K. However, $T_c$ decreases with the increase of Fe content for samples with $x \geqslant 0.25$. No $T_c$ above 40 K has been observed in these Rb-series samples. This Fe-content dependent behavior of $T_c$ is slight different from a $K_xFe_ySe_2$ system, in which $T_c$ remains constant with the increase of Fe content, and two superconducting transitions have been observed in the samples with higher Fe content [14]. This should be caused by the difference of effective Fe content in these two series of samples from ICP data: when the nominal composition of Fe is higher than 1.8 ($x = 0.2$), the effective Fe content in Rb samples increases continuously with the increasing of $x$ while it remains constant in K samples [14]. In addition, we would like to point out that as temperature decreases, the resistivity in the normal state first increases, then peaks, and finally decreases, as shown in figures 2(b) and (c). This kind of hump-like behavior is also commonly observed in other $A_xFe_ySe_2$ superconductors [1–3]. Figure 2(d) shows the resistivity at low temperature for clearly illustrating the superconducting transition.

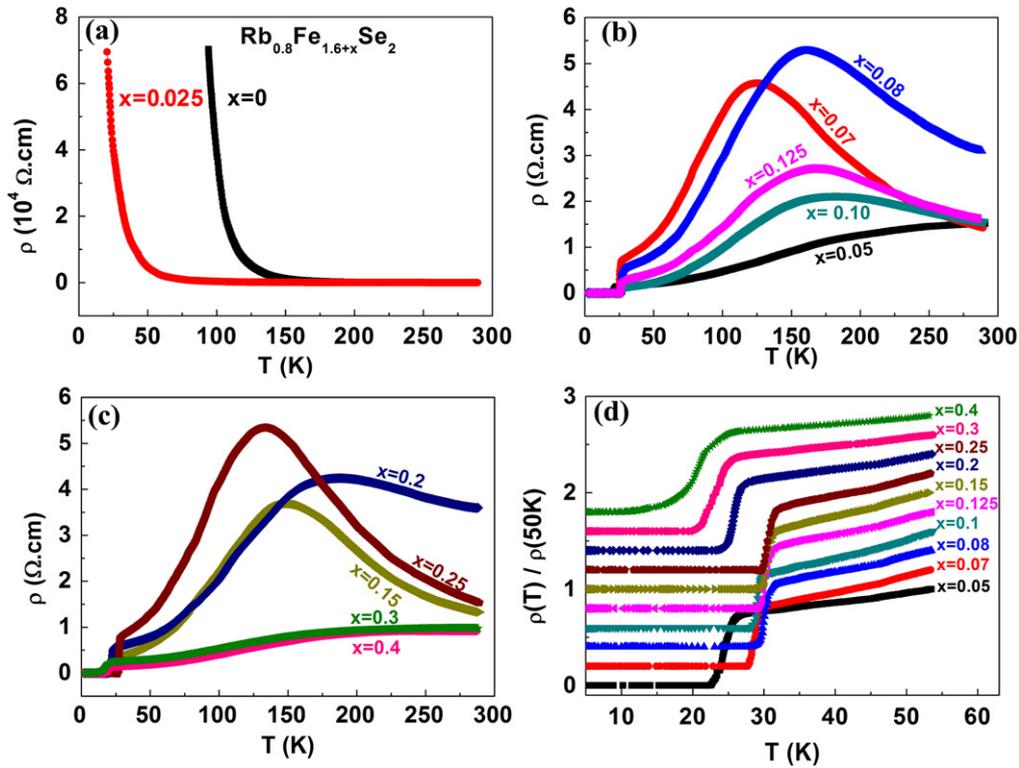

**Figure 2.** (a) Temperature dependence of electric resistivity for $Rb_{0.8}Fe_{1.6+x}Se_2$ samples with $x = 0$ and 0.025. (b)–(c) Temperature dependence of electric resistivity for superconducting samples ($0.05 \leqslant x \leqslant 0.4$). (d) Temperature dependence of electric resistivity for superconducting samples at low temperature, all curves except for $x = 0.05$ are offset by 0.2 for clarity.

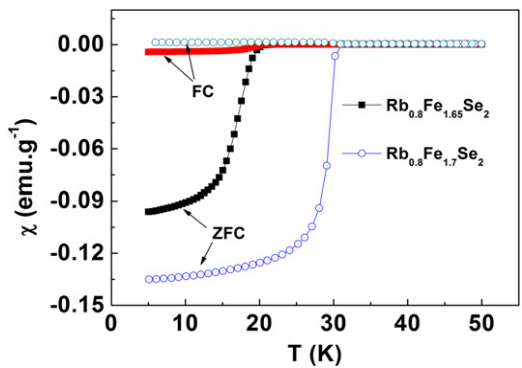

**Figure 3.** Temperature dependence of ZFC and FC magnetic susceptibility of $Rb_{0.8}Fe_{1.6+x}Se_2$ single crystals with $x = 0.05$ and 0.1.

Figure 3 shows the temperature dependence of dc magnetic susceptibility: both zero-field-cooled (ZFC) and field-cooled (FC) for samples of $Rb_{0.8}Fe_{1.65}Se_2$ ($x = 0.05$) and $Rb_{0.8}Fe_{1.7}Se_2$ ($x = 0.1$) measured under applied external magnetic field of 10 Oe. Diamagnetic behaviors are observed below 20 K and 31 K, respectively, which is coincident with resistivity measurement. The applied field was parallel to the $a$–$b$ plane during the measurement; we ignore the demagnetization effect factor since the dimensions of the sample satisfy $a \sim b \gg c$. And the superconducting volume fractions have been estimated as high as 90% at 5 K, which indicates the bulk superconductivity and good quality of our samples.

Based on the $T_c$ inferred from the transport property measurement, the electronic phase diagram has been established in figure 4. $T_c$ increases with the increasing of Fe content when $x \leqslant 0.1$. Here $T_c^{onset}$ is determined from the onset temperature of the downturn where the superconducting transition happens, while $T_c^{zero}$ is the temperature where the resistivity reaches 0, and a quiet flat $T_c$ dome ranging from $x = 0.1$ to 0.2 can be observed; then $T_c$ decreases with the continued increase of Fe content. This superconducting dome is the first observed in this kind of FeSe-122 system. On the other hand, superconducting $T_c$ in $K_{0.8}Fe_{1.6+x}Se_2$ superconductors often shows a rapid increase and reaches the maximum of 31 K, and no superconducting dome has been observed [14]. We can also clearly see that the $T_{hump}$, at which the resistivity reaches its peak, varied with Fe content and shows an 'M' shape. However, not every superconducting sample shows this hump—no hump was observed in the $x = 0.05$ sample and $x \geqslant 0.3$ samples. We hope this electronic phase diagram provides some useful information for understanding the mechanism of superconductivity in this kind of Fe-based superconductor. For example, the superconductivity evolved from the semiconducting samples with the increasing of the Fe content: is the hump related to the presence of the superconductivity or not?

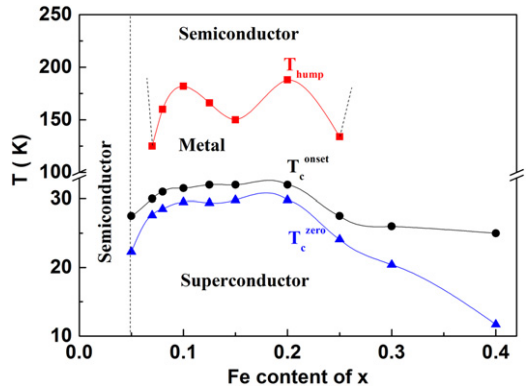

**Figure 4.** Electronic phase diagram of $Rb_{0.8}Fe_{1.6+x}Se_2$. $T_c^{onset}$ is determined from the onset temperature of the downturn where the superconducting transition happens, $T_c^{zero}$ is the temperature where the resistivity reaches 0, and $T_{hump}$ is the temperature where resistivity reaches its peak.

Our dome-like electronic phase diagram looks very similar to that in a $Fe_{1.01}Se$ superconductor at high pressure, where the highest $T_c$ of about 37 K was observed at a pressure of 8.9 GPa [17]. The similarities in both systems are: (a) the highest $T_c$ is more than 30 K; (b) the phase diagrams in both systems do not have any regions of spin-density wave anomaly, which is very common in FeAs-based superconductors [18, 19]. However, unlike the case of $Fe_{1.01}Se$, our $Rb_{0.8}Fe_{1.6+x}Se_2$ samples usually have a hump in the temperature dependent resistivity curves, which occurred from 100–200 K, varied with Fe content, while no such hump was observed in $Fe_{1.01}Se$.

We now focus on discussing the structure and the nature of the phase-separated state in $Rb_{0.8}Fe_{1.6+x}Se_2$ materials. In order to better understand the structural features in this system, we performed an extensive structural characterization by means of selected-area electron diffraction (SAED). Figure 5 shows the electron diffraction patterns taken along the [001] zone-axis direction of the $Rb_{0.8}Fe_{1.6}Se_2$ parent sample and the $Rb_{0.8}Fe_{1.7}Se_2$ superconducting sample. The main diffraction spots with relatively strong intensity can be well indexed by the known tetragonal structure consistent with the XRD results [6]. On the other hand, the most striking structural features revealed in our TEM characterization were the appearance of a series of superlattice spots following the main diffraction spots. These structural features are very similar to that in $K_xFe_ySe_2$ materials. A set of fivefold satellite spots are clearly visible in the $a^* - b^*$ plane of reciprocal space and can be characterized by a unique modulation wave vector $q_1 = 1/5(3a^* + b^*)$, as illustrated in figure 5(a). This superstructure can be interpreted by the Fe-vacancy ordering in the Fe-Se layer [6]. Sometimes, two sets of superstructure reflections appear around each basic Bragg spot, as shown in figure 5(b), in which two sets of superstructure spots are indicated by $q_1$. These two sets of superstructure spots are considered to originate from the domains where the superstructure vectors are twinning-related with respect to one another. This kind of twin domain was also previously observed in K-Fe-Se samples [6, 14].

The coexistence of structural phases has been observed in the $Rb_{0.8}Fe_{1.7}Se_2$ superconducting sample, as shown in figure 5(c). The modulations can be characterized as $q_1 = 1/5[3a^* + b^*]$, the antiferromagnetic $Rb_{0.8}Fe_{1.6}Se_2$ phase, and $q_2 = 1/2[a^* + b^*]$, the superconducting phase, respectively [9, 14, 20, 21]. This structural feature is the same as we observed in superconducting $K_xFe_ySe_2$ samples [14], and the $q_2$ phase can only be observed in the superconducting samples. During our TEM observations, we found that there are more areas with a pure twofold modulation structure in the Rb samples than in the K samples, as indicated in figure 5(d).

A clear view of the phase-separation nature and the related structural characterizations were performed on $Rb_{0.8}Fe_{1.6+x}Se_2$ single crystals by using SEM. The most remarkable microstructural feature revealed by our observations is the appearance of stripe patterns in all superconducting samples, as clearly shown in figure 6. It is recognizable that the density of these dark-contrast stripes increases progressively with the increase of Fe content. This stripe pattern can only be observed in the superconducting samples, which is the same as in K-series samples [14]. In $K_xFe_ySe_2$, it is believed that the appearance of superconductivity in the present system is essentially in correlation with the emergence of these stripe patterns [14]. In general, these stripe patterns consist of Fe-rich crystalline particles and go along the [110] and [1–10] directions. These two directions are actually twinning-related with a 90° rotation with respect to one another, which is the same as in K-Fe-Se samples [14].

In fact, the phase separation in the $Rb_xFe_ySe_2$ system was reported by several groups [16, 21, 22]: intrinsic crystal phase-separation was found in $Rb_xFe_ySe_2$ by synchrotron x-ray and high resolution neutron diffraction; and the minority phase without Fe ordering structure transformed to the main phase with Fe ordering structure at temperature above 475 K [21]. This is similar to the phase transition observed in $K_2Fe_4Se_5$ by TEM [15], where the transition temperature was found to be about 570 K, about 100 K higher than that in the Rb system. The presence of magnetic and nonmagnetic specimens in superconducting $Rb_xFe_ySe_2$ samples found by $^{57}Fe$ Mössbauer studies also provided convincing evidence of phase separation [22].

In summary, we have investigated the superconductivity and microstructure of $Rb_{0.8}Fe_{1.6+x}Se_2$ crystals. The maximum $T_c \approx 32$ K has been obtained in samples with $0.1 \leqslant x \leqslant 0.2$. A superconducting dome in the electronic phase diagram has been observed. Microstructure studies show that the coexistence of two ordered states commonly appears in superconducting samples, which are characterized respectively by the modulations of $q_1 = 1/5[3a^* + b^*]$ for the antiferromagnetic $Rb_{0.8}Fe_{1.6}Se_2$ phase and $q_2 = 1/2[a^* + b^*]$ for the superconducting phase. The phase separation nature and stripe patterns could evidently change with Fe content in the $Rb_{0.8}Fe_{1.6+x}Se_2$ samples; this microstructural feature could play a critical role in the stability of the superconducting phase in $Rb_{0.8}Fe_{1.6+x}Se_2$ materials [23].

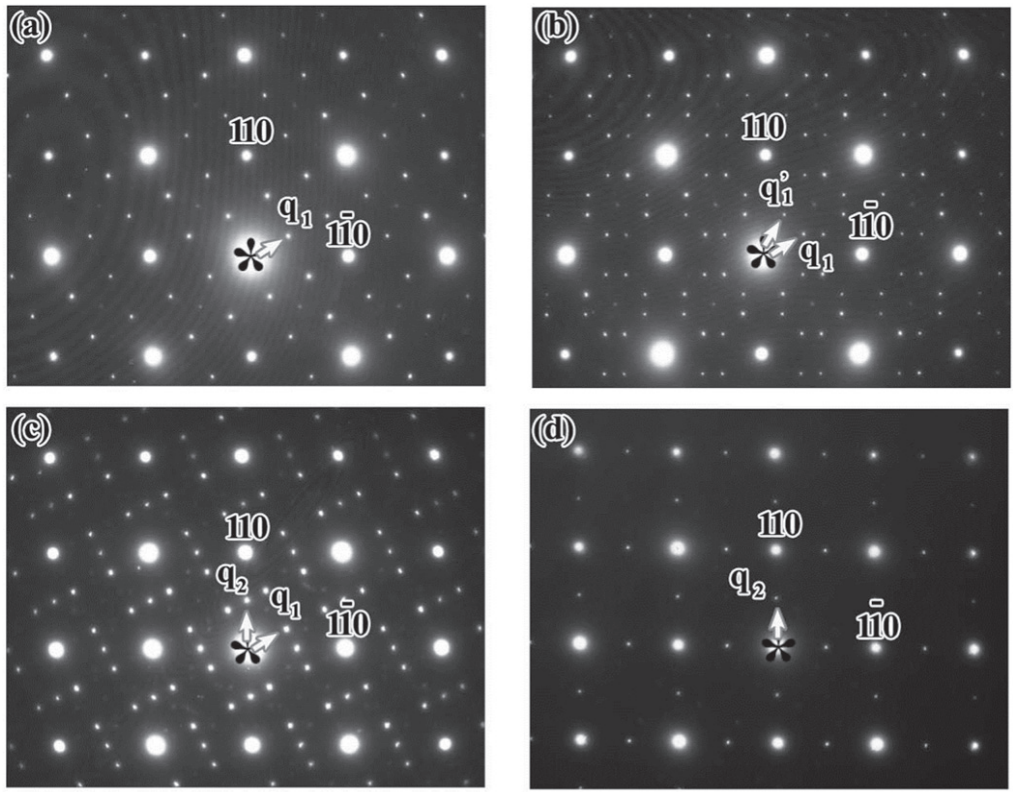

**Figure 5.** Electron diffraction patterns of $Rb_{0.8}Fe_{1.6+x}Se_2$ taken along the [001] zone-axis directions: (a)–(b) taken from the $x=0$ sample, illustrating the superstructure and twinned structure; (c)–(d) taken from the $x=0.08$ sample, illustrating the coexistence of **q1** and **q2** modulations in the phase-separated state.

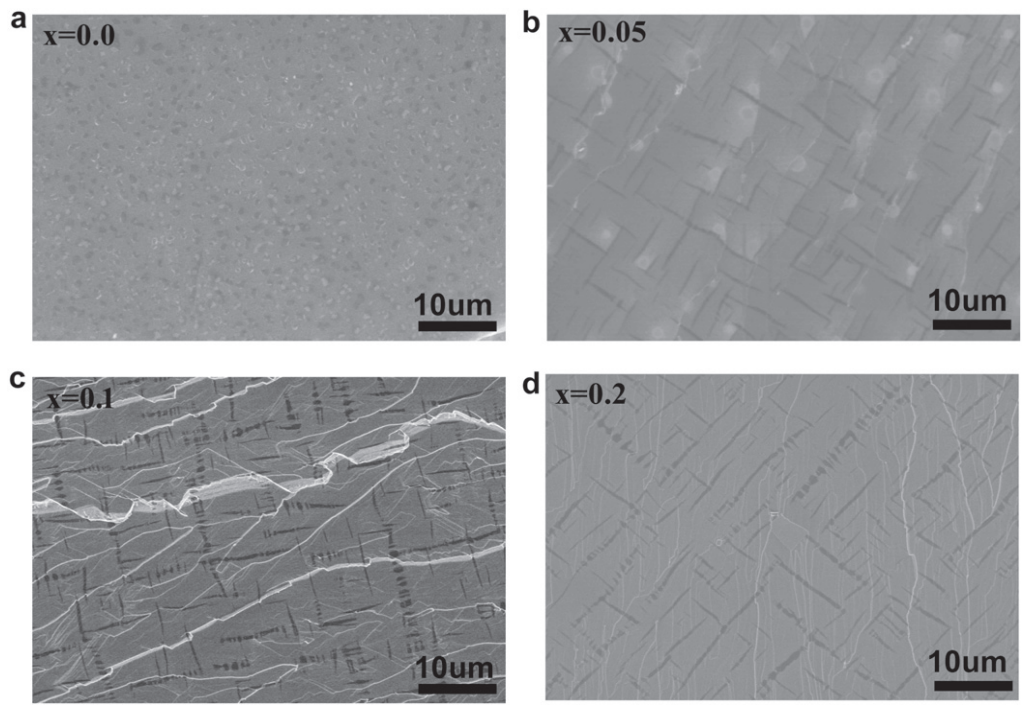

**Figure 6.** SEM images of $a$–$b$ plane reveal the phase-separated state in $Rb_{0.8}Fe_{1.6+x}Se_2$ superconducting samples with (a) $x=0$, (b) 0.05, (c) 0.1 and (d) 0.2, respectively. Stripe density increases visibly with the increase of Fe concentration.

## Acknowledgments

This work was supported by the National Basic Research Program of China 973 Program (grant nos. 2011CB921703, 2015CB921300, 2011CBA00101, 2010CB923002 and 2012CB821404), the Natural Science Foundation of China (grant nos. 11474323, 51272277, 91221102, 11190022 and 11274368) and the Strategic Priority Research Program of the Chinese Academy of Sciences (grant no. XDB07020000).

## References

[1] Guo J G, Jin S F, Wang G, Wang S C, Zhu K X, Zhou T T, He M and Chen X L 2010 *Phys. Rev.* B **82** 180520
[2] Wang D M, He J B, Xia T L and Chen G F 2011 *Phys. Rev.* B **83** 132502
[3] Fang M H, Wang H D, Dong C H, Li Z J, Feng C M, Chen J and Yuan H Q 2011 *EPL* **94** 27009
[4] Li C H, Shen B, Han F, Zhu X Y and Wen H H 2011 *Phys. Rev.* B **83** 184521
[5] Wang A F *et al* 2011 *Phys. Rev.* B **83** 060512
[6] Wang Z, Song Y J, Shi H L, Wang Z W, Chen Z, Tian H F, Chen G F, Guo J G, Yang H X and Li J Q 2011 *Phys. Rev.* B **83** 140505 (R)
[7] Yan X W, Gao M, Lu Z Y and Xiang T 2011 *Phys. Rev. Lett.* **106** 087005
[8] Bao W, Huang Q, Chen G F, Green M A, Wang D M, He J B, Wang X Q and Qiu Y 2011 *Chin. Phys. Lett.* **28** 086104
[9] Torchetti D A, Fu M, Christensen D C, Nelson K J, Imai T, Lei H C and Petrovic C 2011 *Phys. Rev.* B **83** 104508
[10] Yuan R H, Dong T, Chen G F, He J B, Wang D M and Wang N L 2012 *Sci. Rep.* **2** 221
[11] Zhang Y *et al* 2011 *Nat. Mater.* **10** 273
[12] Qian T *et al* 2011 *Phys. Rev. Lett.* **106** 187001
[13] Mou D *et al* 2011 *Phys. Rev. Lett.* **106** 107001
[14] Wang Z W, Wang Z, Song Y J, Ma C, Cai Y, Chen Z, Tian H F, Yang H X, Chen G F and Li J Q 2012 *J. Phys. Chem.* C **116** 17847
[15] Song Y J, Wang Z, Wang Z W, Shi H L, Chen Z, Tian H F, Chen G F, Yang H X and Li J Q 2011 *EPL* **95** 37007
[16] Ricci A *et al* 2011 *Supercond. Sci. Technol.* **24** 082002
[17] Medvedev S *et al* 2009 *Nat. Mater.* **8** 630
[18] Hess C, Kondrat A, Narduzzo A, Hamann-Borrero J E, Klingeler R, Werner J, Behr G and Büchner B 2009 *EPL* **87** 17005
[19] Luetkens H *et al* 2009 *Nat. Mater.* **8** 305
[20] Charnukha A *et al* 2012 *Phys. Rev. Lett.* **109** 017003
[21] Pomjakushin V Y, Pomjakushina E V, Krzton-Maziopa A, Conder K, Chernyshov D, Svitlyk V and Bosak A 2012 *J. Phys.: Condens. Matter* **24** 435701
[22] Ksenofontov V, Wortmann G, Medvedev S A, Tsurkan V, Deisenhofer J, Loidl A and Felser C 2011 *Phys. Rev.* B **84** 180508 (R)
[23] Wang Z, Cai Y, Wang Z W, Sun Z A, Yang H X, Tian H F, Ma C, Zhang B and Li J Q 2013 *EPL* **102** 37010